\documentclass[a4paper,preprint]{revtex4}
\usepackage{epsf}
\usepackage[dvips]{graphicx}
\usepackage{graphics}

\begin{document}
\title{New Results for the Nonlocal Kardar-Parisi-Zhang Equation}
\author{Eytan Katzav}
\email{eytak@post.tau.ac.il} \affiliation {School of Physics and
Astronomy, Raymond and Beverly Sackler Faculty of Exact Sciences,
Tel Aviv University, Tel Aviv 69978, Israel}

\begin{abstract}
In this paper various predictions for the scaling exponents of the
Nonlocal Kardar-Parisi-Zhang (NKPZ) equation are discussed. I use
the Self-Consistent Expansion (SCE), and obtain results that are
quite different from result obtained in the past, using Dynamic
Renormalization Group analysis (DRG), a Scaling Approach (SA) and
a self-consistent Mode Coupling approach (MC). It is shown that
the results obtained using SCE recover an exact result for a
subfamily of the NKPZ models in one dimension, while all the other
methods fail to do so. It is also shown that the SCE result is the
only one that is compatible with simple observations on the
dependence of the dynamic exponent $z$ in the NKPZ model on the
exponent $\rho$ characterizing the decay of the nonlinear
interaction. The reasons for the failure of other methods to deal
with NKPZ are also discussed.
\end{abstract}

\maketitle

\section{Introduction}
The field of disorderly surface growth has received much attention
during the last two decades. A multitude of different phenomena
such as fluid flow in porous media, propagation of flame fronts,
flux lines in superconductors not to mention deposition processes,
bacterial growth and "DNA walk" \cite{barabasi95} are all said to
be related to the famous Kardar-Parisi-Zhang (KPZ) equation
\cite{kpz86}. Therefore, further understanding of the behavior of
the KPZ equation has a broad interest in the fields of
nonequilibrium dynamics and in disordered systems.

The KPZ equation for a growing surface is
\begin{equation}
\frac{{\partial h\left( {\vec r,t} \right)}}{{\partial t}} = \nu
\nabla ^2 h + \frac{\lambda }{2}\left( {\nabla h} \right)^2  +
\eta \left( {\vec r,t} \right)
 \label{1},
\end{equation}
This equation describes the height fluctuations of a d-dimensional
interface, where $h\left( {\vec r,t} \right)$ is the height at
$\vec r$ measured relative to its spatial average, $\nu $ is the
diffusion constant and $\eta \left( {\vec r,t} \right)$ is the
fluctuation of the rate of deposition. The noise $\eta \left(
{\vec r,t} \right)$ has a zero mean $\left( {\left\langle {\eta
\left( {\vec r,t} \right)} \right\rangle  = 0} \right)$ and it
satisfies
\begin{equation}
\left\langle {\eta \left( {\vec r,t} \right)\eta \left( {\vec
r',t'} \right)} \right\rangle  = 2D_0 \delta ^d \left( {\vec r -
\vec r'} \right)\delta \left( {t - t'} \right)
 \label{2},
\end{equation}
(white noise) where $d$ is the substrate dimension and $D_0$
specifies the noise amplitude.

Together with the great success of the KPZ equation to describe
many growth models and phenomena, there has been a growing pool of
data that is not well described by KPZ, and triggered further
research. One of the first classes that belongs to this non KPZ
behavior is the well known Molecular-Beam-Epitaxy (MBE) class
(sometimes also called the conserved KPZ equation)
\cite{wolf90}-\cite{sun93}. This class is distinct from KPZ in
having surface diffusion as the basic relaxation mechanism.
However, the modified behavior introduced by the MBE equation is
not at all sufficient to account for all the rich non KPZ
experimental data in the field.

A different line of research suggested that the basic growth
equation should not be changed. Instead, the white noise that
appears in the original KPZ equation should be correlated - either
temporally or spatially \cite{medina89}-\cite{katzav99}. This
approach was indeed quite successful. However, it still failed to
give a good account for all the measured scaling exponents.

Recently, some researchers suggested that incorporating the
long-range nature of interactions in the growing surface is
necessary for a proper description of many systems such as colloid
systems \cite{Lei96}, paper-burning experiments \cite{Zhang92} or
protein deposition kinetics \cite{Ramsden93}. Following this basic
intuition Mukherji and Bhattacharjee \cite{Mukh97} developed a
Langevin-type equation with a nonlocal non-linearity, thus going
beyond the KPZ local description of interactions. They studied
originally the white noise case that was later generalized to
spatially correlated noise by A. Kr. Chattapohadhyay
\cite{chat99}. To be more specific, the equation they studied was
\begin{equation}
\frac{{\partial h\left( {\vec r,t} \right)}}{{\partial t}} = \nu
\nabla ^2 h\left( {\vec r,t} \right) + \frac{1}{2}\int {dr'g\left(
{\vec r'} \right)\nabla h\left( {\vec r + \vec r',t} \right) \cdot
\nabla h\left( {\vec r - \vec r',t} \right)}  + \eta \left( {\vec
r,t} \right)
\label{3},
\end{equation}
where the kernel $g\left( {\vec r} \right)$ represents the
long-range interactions. They take $g\left( {\vec r} \right)$ with
a short-range part $\lambda _0 \delta ^d \left( {\vec r} \right)$
and a long-range part $ \sim \lambda _\rho  r^{\rho  - d} $, or
more precisely in Fourier space, $\hat g\left( q \right) = \lambda
_0  + \lambda _\rho  q^{ - \rho } $. The noise has again a zero
mean, but it is allowed to have spatially long range correlations,
characterized by its second moment
\begin{equation}
\left\langle {\eta \left( {\vec r,t} \right)\eta \left( {\vec
r',t} \right)} \right\rangle  = 2D_0 \left| {\vec r - \vec r'}
\right|^{2\sigma  - d} \delta \left( {t - t'} \right)
\label{4},
\end{equation}
where the case of white noise is restored in the limit $\sigma  =
0$ \cite{Mukh97}.

In the next section I will present current theoretical predictions
for the scaling exponents of the NKPZ model. It will become
evident that all the previous results are inconsistent with an
exact one-dimensional result obtained in the past. Then, in
section III the Self-Consistent Expansion (SCE) approach will be
applied to this model. Interestingly, this method yields results
that are consistent with the exact one-dimensional result. Then,
results for higher dimensions as well as for general $\rho$'s are
presented. Eventually, a discussion of the reasons for the failure
of the various theoretical methods is given in section IV, and
conclusions are drawn in section V.

\section{Previous Results}
Both papers \cite{Mukh97}-\cite{chat99} investigated this problem
using Dynamic Renormalization Group (DRG) analysis, and derived a
very complex zoo of universality classes (ref. \cite{chat99} also
applied a mode-coupling approach confirming some of the results of
DRG). Let us not get into all those details, but rather focus on
the strong coupling solution (in the KPZ sense \cite{barabasi95})
suggested by both papers, namely
\begin{equation}
z_{DRG} = 2 + \frac{{\left( {d - 2 - 2\rho } \right)\left( {d - 2
- 3\rho } \right)}}{{\left( {3 + 2^{ - \rho } } \right)d - 6 -
9\rho }} \label{5},
\end{equation}
where $z$ is the dynamic exponent. To complete the picture, the
roughness exponent, $\alpha $, can be calculated using the
modified scaling relation $\alpha  + z = 2 - \rho $ (which
actually comes from the famous Galilean invariance). Notice that
this result does not depend explicitly on $\sigma$ as long as
$\sigma > 0$.

On the basis of the DRG result (eq. \ref{5}) Mukherji and
Bhattacharjee tried to explain some non KPZ experimental results
in one dimension, namely a roughness exponent of $\alpha  = 0.71$
given in refs. \cite{Lei96}-\cite{Zhang92} that implies, according
to their DRG calculation, $\rho=- 0.12$, and requires $\lambda _0
= 0$. As already appreciated by the authors, this result raises
some doubts on physical grounds, whether a true $\lambda _0  = 0$
is seen in experiment, and where should a negative $\rho $, which
implies anticorrelations \cite{Villain} rather than correlations
that cause the interaction, come from.

This finding is obviously disturbing because if DRG yields
accurate results then the NKPZ equation cannot be relevant to the
physical processes discussed above. As will be shown later, the
DRG result is far from being accurate, so that the above argument
against the relevance of the NKPZ equation is not valid. The
relevance of the NKPZ equation has been doubted, however, for
other reasons too, but this will not be discussed here.

The equation is interesting by itself and generalizes the
traditional KPZ prototype of nonlinear stochastic field equations
that are so abundant in the description of natural phenomena.
Therefore, reliable methods of solution are of great importance.
DRG is such a general method, but unfortunately the DRG results
look suspicious for three reasons:

First, the expression for the dynamic exponent (eq. (\ref {5}))
reduces to the well-known result of Medina et. al. \cite{medina89}
$z = 2 + \frac{{\left( {d - 2} \right)^2 }}{{4d - 6}}$ in the
limit of $\rho  = 0$ (i.e. the limit of local KPZ) - a result that
by itself is meaningful only for the case of $d = 1$. It is a
well-known fact that DRG cannot give the strong coupling solution
of the local KPZ equation for $d > 1$ \cite{Wiese98}. This may
suggest that the DRG result is not correct also for the nonlocal
case, especially for $d>1$. Furthermore, the mode coupling
approach that is known to be more appropriate than DRG when
dealing with the standard KPZ problem \cite{chat98, BC93} yielded
a strong coupling result that is different from eq. (\ref{5})
\cite{chat99}. Yet, it should be mentioned that the mode coupling
approach \cite{chat99} (at least using the simplest ansatz) was
not able to produce a non power counting solution, as obtained in
the local KPZ case \cite{BC93}.

Second, it is easily seen that the explicit expression for the
dynamical exponent (given in eq. (\ref{5}) above) has a
singularity for $\rho $'s that satisfy $\left( {3 + 2^{ - \rho _s
} } \right)d - 6 - 9\rho _s  = 0$ (for example in one
dimension$\rho _s =  - 0.205$). It is not clear why should such a
singularity appear at all and if it appears, then why should it
take this specific value.

Third, from general considerations one should expect the following
scenario: as $\rho $ becomes more and more negative, the
interaction that couples the gradients is enormously reduced with
distance (remember that the long-range part of the kernel scales
as $g\left( {\vec r} \right) \sim \lambda _\rho  r^{\rho  - d}$).
Therefore, one should expect that in this region, the nonlinear
term becomes irrelevant, and the scaling exponents of the linear
theory (also known as the Edwards-Wilkinson (EW) theory \cite{EW})
should be recovered. On the other side, as $\rho $ becomes larger
and larger, the interaction becomes more and more relevant, and
one might expect that as a consequence faster relaxations should
appear in the system (faster relaxations imply smaller values of
the dynamical exponent $z$). Therefore, it is reasonable that a
non-increasing behavior of the dynamical exponent ($z$) as a
function of $\rho $ will be seen. It is not difficult to see that
the expression for the dynamical exponent (given by eq. (\ref{5}))
does not follow this reasoning.

Lately, an exact result for the NKPZ model was found
\cite{NKPZ02}. It turns out that the Fokker-Planck equation
associated with the Langevin-form of the NKPZ model (i.e eq.
(\ref{3}) above) can be solved exactly for a specific sub-family
of models in one dimension. More specifically, a Gaussian
steady-state solution was found for the case $\lambda_0=0$, $\rho
= 2\sigma $ when $d = 1$. This exact solution yields the following
dynamic exponent
\begin{equation}
z_{exact} = \frac{{3 - 3\rho }}{2} \label{6}.
\end{equation}
It is easy to see that this result reduces to the well know
local-KPZ (that corresponds to $\rho  = 0$ ) result, $z = {3
\mathord{\left/ {\vphantom {3 2}} \right.
\kern-\nulldelimiterspace} 2}$, in one dimension.

This exact result is not compatible with the DRG result - either
quantitatively nor qualitatively. As mentioned above, DRG is
usually considered relatively reliable in one dimension. However,
this exact result indicates DRG's shortcoming already in one
dimension. The inevitable conclusion is that in order to gain
insight into the behavior of the NKPZ model in any dimension, and
for any value of $\rho $ (the nonlocal parameter) and $\sigma $
(the long-range noise parameter) it is necessary to employ more
reliable methods.

Actually, two methods that proved useful in the context of the
local-KPZ problem were applied recently to the NKPZ problem as
well. The first method, is a Flory-type scaling approach that was
originally proposed by Hentschel and Family \cite{Hen91} and
generalized lately by Tang and Ma \cite{Tang01} to the nonlocal
case. The strong-coupling dynamical exponent obtained by using
this Scaling Approach (SA) is
\begin{equation}
z_{SA} = \frac{{\left( {2 - \rho } \right)\left( {2 + d - 2\sigma
} \right)}}{{3 + d - 2\sigma }} \label{7}.
\end{equation}

The second method applied to the NKPZ equation is an improved
version of the mode-coupling approach. More precisely, it is a
solution of the mode-coupling equations using an improved ansatz
that describes correctly the asymptotic behavior of the solution.
This approach was first proposed by Colaiori and Moore
\cite{Moore01}, and applied by Hu and Tang \cite{Hu02} to the
nonlocal case. The strong-coupling exponents using this method are
obtained by a numerical solution of the following set of
transcendental equations
\begin{eqnarray}
\frac{{PAS'_d }}{B} = \frac{{d\left( {2 - z} \right)}}{{\pi z^2
}}\frac{{2^{2\rho } }}{{BI\left( {B,z,\rho } \right)}} \nonumber\\
\nonumber\\  \frac{{PAS'_d }}{{B^2 }} = \frac{{\beta \left( {\beta
- z} \right)}}{{\pi z^2 }}\frac{{2^{{{\left( {2z - \beta }
\right)} \mathord{\left/
 {\vphantom {{\left( {2z - \beta } \right)} \beta }} \right.
 \kern-\nulldelimiterspace} \beta } + 2\rho } }}{{{{\Gamma \left( {{\textstyle{{2z - \beta } \over \beta }}} \right)} \mathord{\left/
 {\vphantom {{\Gamma \left( {{\textstyle{{2z - \beta } \over \beta }}} \right)} \beta }} \right.
 \kern-\nulldelimiterspace} \beta }}}
  \label{8}
\end{eqnarray}
where $S'_d  = \int_0^\pi  {\sin ^{d - 2} \theta d\theta } $,
$\Gamma \left( u \right)$ is Euler's gamma function, $\beta  = d +
4 - 2z - 2\rho $, $B = \frac{1}{{2\left( {2 - z} \right)}}$ and
$I\left( {B,z,\rho } \right) = \int_0^\infty  {\left( {1 - \rho  -
2s^2 } \right)s^{2z - 3} \exp \left( { - B^{{\beta \mathord{\left/
{\vphantom {\beta  z}} \right. \kern-\nulldelimiterspace} z}}
s^\beta   - s^2 } \right)ds} $. Such a numerical solution for $z$
as a function of $\rho $ for $d = 1,2,3$ is presented in FIG
\ref{exps-c} of ref. \cite{Hu02}.

While these two results look reasonable, in the sense that both do
not develop any singularity as a function of $\rho $, and at least
the first one (given by eq. (\ref{7})) describes a non-increasing
dynamical exponent, it turns out that both results are still
incompatible with the exact result of $z = {{(3 - 3\rho) }
\mathord{\left/ {\vphantom {{(3 - 3\rho } 2}} \right.
\kern-\nulldelimiterspace} 2}$ (given by eq. (\ref{6})). In
addition, the mode-coupling solution has a peak for $d > 1$
meaning that the dynamic exponent $z$ is not monotonous as a
function of $\rho$ that is expected from general considerations.
Therefore, there is still a need for a reliable method to tackle
this problem. The reasons given above motivated the analysis
presented in this paper.

\section{The Self Consistent Expansion}
In this paper I apply a method developed by Schwartz and Edwards
\cite{SE92,SE98,katzav99} (also known as the
Self-Consistent-Expansion (SCE) approach). This method has been
previously applied successfully to the KPZ equation. The method
gained much credit by being able to give a sensible prediction for
the KPZ critical exponents in the strong coupling phase for $d>1$,
where, as previously mentioned, many Renormalization-Group (RG)
approaches failed (as well as DRG of course). It also produces the
exact one-dimensional result. I will show that in contrast to DRG,
the experimental result $\alpha = 0.71$ is accounted for, when
using the SCE, by a positive $\rho $. Furthermore, I will show
that the SCE method yields the exact result, for the specific
sub-family of models that can be solved exactly in one dimension -
where other methods fail. In addition, solutions for any $d,\rho $
and $\sigma $ are obtained. These solutions are consistent with
the expected qualitative scenario presented above.

As will be seen immediately, another remarkable advantage of the
SCE method is the minor changes needed in order to generalize the
result of local KPZ with uncorrelated noise to include nonlocal
interactions as well as spatially correlations in the noise. The
above implies a second important motivation for this paper, namely
a demonstration of the robustness of the SCE method as well as its
mathematical coherence and consistency.

The SCE method is based on going over from the Fourier transform
of the equation in Langevin form to a Fokker-Planck form and
constructing a self-consistent expansion of the distribution of
the field concerned.

The expansion is formulated in terms of $\phi _q $ and $\omega _q
$, where $\phi _q $ is the two-point function in momentum space,
defined by $\phi _q  = \left\langle {h_q h_{ - q} } \right\rangle
_S $, (the subscript S denotes steady state averaging), and
$\omega _q $ is the characteristic frequency associated with each
mode.

It is generally expected that for small enough $q$, $\phi _q $ and
$\omega _q $ are power laws in $q$,

\begin{equation}
\phi _q  = Aq^{ - \Gamma } \label{9}
\end{equation}
and
\begin{equation}
\omega _q  = Bq^z
\label{10},
\end{equation}
where $z$ is just the dynamic exponent, and the exponent $\Gamma$
is related to the roughness exponent $\alpha $ by

\begin{equation}
 \alpha  = \frac{{\Gamma  - d}}{2}
\label{11}.
\end{equation}

The method produces, to second order in this expansion, two
nonlinear coupled integral equations in $\phi _q $ and $\omega _q
$, that can be solved exactly in the asymptotic limit to yield the
required scaling exponents governing the steady state behavior and
the time evolution.

In the following I will find the critical exponents of the NKPZ
equation with spatially correlated noise. I consider the NKPZ
equation (eq. (\ref{3})) in Fourier components
\begin{equation}
\frac{{\partial h_q }}{{\partial t}} =  - \nu _q h_q  -
\frac{1}{{\sqrt \Omega  }}\sum\limits_{\ell ,m} {M_{q\ell m}
h_\ell  h_m }  + \eta _q
\label{12},
\end{equation}
where $\Omega $ is the volume of the system, to be taken
eventually to infinity and $\eta _q$ is the noise term that
satisfies:
\begin{equation}
\left\langle {\eta _q \left( t \right)} \right\rangle  = 0
\label{13},
\end{equation}
\begin{equation}
\left\langle {\eta _q \left( t \right)\eta _{q'} \left( {t'}
\right)} \right\rangle  = 2D_q \delta ^d \left( {\vec q + \vec q'}
\right)\delta \left( {t - t'} \right)
\label{14},
\end{equation}
and $M_{q\ell m}$, $\nu _q$ and $D_q$ are defined by
\begin{eqnarray}
 M_{q\ell m} &=& \frac{{\lambda _\rho  }}{{\sqrt \Omega  }}q^{ - \rho } \left( {\vec \ell  \cdot \vec m} \right)\delta _{q,\ell  + m}  \nonumber\\
 \nu _q  &=& \nu q^2  \nonumber\\
 D_q  &=& D_0 q^{ - 2\sigma }
\label{15}.
\end{eqnarray}

Thus, the NKPZ equation is of the general form discussed in refs.
\cite{SE92, SE98}, where the SCE is derived. Working to second
order in the expansion, this yields the two coupled non-linear
equations
\begin{equation}
D_q  - \nu _q \phi _q  + 2\sum\limits_{\ell ,m} {\frac{{M_{q\ell
m} M_{q\ell m} \phi _\ell  \phi _m }}{{\omega _q  + \omega _\ell +
\omega _m }}}  - 2\sum\limits_{\ell ,m} {\frac{{M_{q\ell m}
M_{\ell mq} \phi _m \phi _q }}{{\omega _q  + \omega _\ell   +
\omega _m }}}  - 2\sum\limits_{\ell ,m} {\frac{{M_{q\ell m}
M_{m\ell q} \phi _\ell  \phi _q }}{{\omega _q  + \omega _\ell   +
\omega _m }}}  = 0
\label{16},
\end{equation}
and
\begin{equation}
\nu _q  - \omega _q  - 2\sum\limits_{\ell ,m} {M_{q\ell m}
\frac{{M_{\ell mq} \phi _m  + M_{m\ell q} \phi _\ell  }}{{\omega
_\ell   + \omega _m }}}  = 0
\label{17},
\end{equation}
where in deriving the last equation I have used the Herring
consistency equation \cite{Herring}. In fact Herring's definition
of $\omega _q $ is one of many possibilities, each leading to a
different consistency equation. But it can be shown, as previously
done in \cite{SE98}, that this does not affect the exponents
(universality).

A detailed solution of equations (\ref{16}) and (\ref{17}) in the
limit of small $q's$ (i.e. large scales) is performed (as in refs.
\cite{katzav99,SE98}), and yields a very rich family of solutions
for any $d,\rho $ and $\sigma $. These solutions are given in
Table~\ref{exp_vals}. However, before getting into all the details
let me focus on the strong coupling non-power counting solution
obtained by SCE. The reason for focusing on this solution is that
it is interesting to compare it with the solutions obtained by
other methods, and with the experimental results. This solution is
determined from the combination of the scaling relation $z  =
\frac{{d + 4 - \Gamma  - 2\rho }}{2}$, obtained from eq.
(\ref{17}), and the transcendental equation$F\left( {\Gamma ,z
,\rho } \right) = 0$, where F is given by
\begin{eqnarray}
 F\left( {\Gamma ,z ,\rho } \right) =  &-& \int {d^d t\frac{{\vec t \cdot \left( {\hat e - \vec t} \right)}}{{t^z   + \left| {\hat e - \vec t} \right|^z   + 1}}\left[ {\left( {\hat e \cdot \vec t} \right)\left| {\hat e - \vec t} \right|^{ - \rho } t^{ - \Gamma }  + \hat e \cdot \left( {\hat e - \vec t} \right)t^{ - \rho } \left| {\hat e - \vec t} \right|^{ - \Gamma } } \right]}  +  \nonumber\\
 \quad \quad \quad \quad \quad \quad  &+& \int {d^d t\frac{{\left[ {\vec t \cdot \left( {\hat e - \vec t} \right)} \right]^2 }}{{t^z   + \left| {\hat e - \vec t} \right|^z   + 1}}t^{ - \Gamma } \left| {\hat e - \vec t} \right|^{ - \Gamma } }
 \label{18},
\end{eqnarray}
and $\hat e$ is a unit vector in an arbitrary direction.

This solution is valid as long as it satisfies the following
conditions: $d < 3\Gamma _0 \left( {d,\rho } \right) - 4 + \min
\left\{ {2\rho  - 4\sigma , - 2\rho } \right\}$ and $d < \Gamma _0
\left( {d,\rho } \right) - 2\left| \rho  \right|$, where $\Gamma
_0 \left( {d,\rho } \right)$ is the numerical solution of the
equation $F\left( {\Gamma ,z ,\rho } \right) = 0$ - when such a
solution exists (how to obtain the conditions above see ref.
\cite{SE98}).

It turns out that for $d = 1$ the equation $F\left( {\Gamma ,z
\left( \Gamma  \right),\rho } \right) = 0$ is exactly solvable for
any $\rho $, and yields $\Gamma  = 2 + \rho $ and $z  = \frac{{3 -
3\rho }}{2}$ (it can be checked immediately by direct
substitution). In this case the three validity conditions read ${{
- 1} \mathord{\left/ {\vphantom {{ - 1} 3}} \right.
\kern-\nulldelimiterspace} 3} < \rho  < 1$ and $\rho  >
\frac{{4\sigma  - 1}}{5}$. By using eq. (\ref{11}) I translate the
results into $\alpha  = \frac{{\rho  + 1}}{2}$ and $z = \frac{{3 -
3\rho }}{2}$. It is straightforward to check that this solution
reduces to the exact KPZ results of $\alpha  = {1 \mathord{\left/
{\vphantom {1 2}} \right. \kern-\nulldelimiterspace} 2}$ and $z =
{3 \mathord{\left/ {\vphantom {3 2}} \right.
\kern-\nulldelimiterspace} 2}$ in the limit of $\rho  = 0$.
Moreover, this solution recovers the exact solution obtained in
ref. \cite{NKPZ02}, and by doing that clearly does better than all
the other results presented above (eqs. (\ref{5}),(\ref{7}) and
(\ref{8})).

It should also be mentioned that for $d \ge 2$ such an exact
solution for $F\left( {\Gamma ,z \left( \Gamma  \right),\rho }
\right) = 0$ as a closed analytical expression cannot be found,
and one has to solve numerically the equation in order to find
$\Gamma _0 \left( {d,\rho } \right)$ (such results are presented
below).

As can be appreciated, the full description of the results given
in TABLE 1 is quite rich and may look confusing. In order to gain
more insight into this seemingly long list of possible phases I
concentrate on the case of white noise $\left( {\sigma  = 0}
\right)$, and I describe all the possible phases (as a function
of $\rho $) for $d = 1,2,3$.

In one dimension, the following scaling exponents are obtained:
\begin{equation}
z  = \left\{ \begin{array}{l}
 2\quad \quad \quad \quad \rho  <  - {\textstyle{1 \over 2}} \\
 \frac{{5 - 2\rho }}{3}\quad \quad  - {\textstyle{1 \over 2}} < \rho  <  - {\textstyle{1 \over 5}} \\
 \frac{{3 - 3\rho }}{2}\quad \quad  - {\textstyle{1 \over 5}} < \rho  < 1 \\
 2 - 2\rho \quad \quad \quad \quad \rho  > 1 \\
 \end{array} \right.\quad \quad and\quad \quad \Gamma  = \left\{ \begin{array}{l}
 2\quad \quad \quad \quad \rho  <  - {\textstyle{1 \over 2}} \\
 \frac{{5 - 2\rho }}{3}\quad \quad  - {\textstyle{1 \over 2}} < \rho  <  - {\textstyle{1 \over 5}} \\
 2 + \rho \quad \;\;\quad  - {\textstyle{1 \over 5}} < \rho  < 1 \\
 2\quad \quad \quad \quad \rho  > 1 \\
 \end{array} \right.
 \label{19}.
\end{equation}

It can be seen that the dynamic exponent obtained using SCE is
continuous, has no singularities and is non-increasing as a
function of $\rho $. In addition, the intuition that the scaling
exponents of the linear theory should appear in the limit of very
small $\rho 's$ is verified and specified, namely for $\rho  \le
- {\textstyle{1 \over 2}}$ the Edwards-Wilkinson (EW) exponents
of $z = 2$ and $\alpha  = {1 \mathord{\left/ {\vphantom {1 2}}
\right. \kern-\nulldelimiterspace} 2}$ are obtained.

An important observation regarding the one-dimensional results is
that for every region of $\rho 's$, well defined values of $z$ and
$\Gamma $ are obtained, so that only one phase is possible when
$\rho $ is specified, and no phase transition as a function of the
dimensionless coupling constant (that is defined by
$g^2=\frac{\lambda^2D}{\nu^3}$) exists.

In two dimensions, the following scaling exponents are obtained:
\begin{equation}
z  = \left\{ \begin{array}{l}
 2\quad \quad \quad \quad \quad \quad \quad \rho  < 0 \\
 2 - 2\rho \quad \quad \quad \quad \quad \,\rho  > 0 \\
 {\textstyle{{2 - \Gamma _0 \left( {2,\rho } \right)} \over 2}} + 2 - \rho \quad \quad \rho _2^\ell   < \rho  < \rho _2^u  \\
 \end{array} \right. \quad and \quad \Gamma  = \left\{ \begin{array}{l}
 2\quad \quad \quad \quad \quad \rho  < 0 \\
 2\quad \quad \quad \quad \quad \rho  > 0 \\
 \Gamma _0 \left( {2,\rho } \right)\quad \quad \rho _2^\ell   < \rho  < \rho _2^u  \\
 \end{array} \right.
\label{20},
\end{equation}
where $\rho _2^\ell$ and $\rho _2^u$ are the lower and upper
bounds on $\rho $ (respectively) for which a solution for the
transcendental equation $F\left( {\Gamma ,z \left( \Gamma
\right),\rho } \right) = 0$ yields a dynamic exponent that does
not exceed $z = 2$ and $z = 2 - 2\rho $. To be more specific,
$\rho _2^\ell   = - 0.395$ and $\rho _2^u  = 0.305$ (the dynamic
exponent is also presented in FIG. 1 (b)).

An important observation regarding the two-dimensional result is
that for values of $\rho $ within the region $\rho _2^\ell   <
\rho  < \rho _2^u$ there are two possible solutions (except for
$\rho=0$). This can be easily seen in FIG. 1 (b) - where two
branches appear in that region. This means that a phase transition
is possible, as a function of the strength of the dimensionless
coupling constant in the theory, between a weak coupling solution
(the upper branch in the figure) and a strong coupling solution
(the lower branch in the figure).

This phenomenon is surprising in view of the known fact that
$d=2$ is the lower critical dimension of the local KPZ theory.
This means that for dimensions higher than $2$ a phase transition
becomes possible in the local problem, but in 2D is only the
strong coupling phase. This fact is indeed respected by the
results presented above, because for $\rho=0$ (that corresponds
to the local theory) such a phase transition is not possible.
However, surprisingly if one takes either positive or negative
$\rho$'s (even slightly higher or lower than zero) such a phase
transition becomes possible. This finding is explained by the
fact that the lower critical dimension is lowered in the presence
of long-range interactions. The key point is that for positive
$\rho$'s ($\rho>0$) the relevant "weak coupling" phase is a phase
with a dynamic exponent of $z=2-2\rho$, and this phase becomes
possible for $d>2-\rho$. In addition, for negative $\rho$'s
($\rho<0$), a different "weak coupling" solution is possible -
this one with a dynamic exponent of $z=2$, and this phase becomes
possible for $d>2+2\rho=2-2|\rho|$. Therefore, once we encounter
a non-zero $\rho$ the lower critical dimension is reduced (for
two different reasons), and one of the weak coupling solutions
becomes possible. In that case, a phase transition can also take
place.

In more than two dimensions, the scaling exponents follow the
scheme:
\begin{equation}
z  = \left\{ \begin{array}{l}
 2\quad \quad \quad \quad \quad \quad \quad \rho  \le 0 \\
 2 - 2\rho \quad \quad \quad \quad \quad \,\rho  > 0 \\
 {\textstyle{{d - \Gamma _0 \left( {d,\rho } \right)} \over 2}} + 2 - \rho \quad \quad \rho _d^\ell   < \rho  < \rho _d^u  \\
 \end{array} \right.\quad and\quad \Gamma  = \left\{ \begin{array}{l}
 2\quad \quad \quad \quad \quad \rho  \le 0 \\
 2\quad \quad \quad \quad \quad \rho  > 0 \\
 \Gamma _0 \left( {d,\rho } \right)\quad \quad \rho _d^\ell   < \rho  < \rho _d^u  \\
 \end{array} \right.
\label{21},
\end{equation}
whenever a solution for $F\left( {\Gamma ,z \left( \Gamma
\right),\rho } \right) = 0$ exists. Notice a difference from the
two-dimensional case in that the weak-coupling exponents are also
possible when $\rho  = 0$. I have performed the detailed
calculation for $d = 3$, and it is shown in FIG. 1 (c) (where
$\rho _3^\ell   =  - 0.016$ and $\rho _3^u  =
{\rm{0}}{\rm{.246}}$). It can be seen that no big quantitative
difference from the two-dimensional case is present.

\section{Discussion}
The results I obtained using the self-consistent expansion (SCE)
are obviously different from those obtained by all the other
methods. First, the SCE is able to reproduce the exact
one-dimensional result, where all the other methods fail. Second,
the SCE results are compatible with the expected behavior of the
dynamic exponent for $\rho  \to  - \infty $ (i.e. recovery of the
EW exponents) as well as with the expected non-increasing behavior
of the dynamic exponent as a function of $\rho $. None of the
other results mentioned before recover the EW behavior for large
negative $\rho 's$, and only one such result shows a
non-increasing behavior of the dynamic exponent as a function of
$\rho $ - namely the result presented in ref. \cite{Tang01}.

At this point, it might be interesting to return to the
experimental result of refs. \cite{Lei96}-\cite{Zhang92} that were
discussed previously, and to discuss them in view of the new
results derived here. It turns out that the SCE method yields
$\rho  = 0.42$ $\left( {\rho  = 2\alpha  - 1} \right)$ for the
case of $\alpha  = 0.71$ (at least if the noise is white, or more
precisely if the exponent that describes the decay of spatial
correlations $\sigma$ is not large). This result is physically
more reasonable than the result $\rho =  - 0.12$ suggested in ref.
\cite{Mukh97}. In addition, since $\rho
> 0$, it does not require to impose the strict requirement
$\lambda _0 = 0$, that is also problematic physically.

It remains a mystery why all the methods that were mentioned at
the beginning of the paper failed to reproduce the exact
one-dimensional result, while they all yield the exact result for
the local KPZ case ($\rho  = 0$). The answer to this puzzle lies,
in my opinion, in the fact that all those methods fail to notice
that the nonlocal nonlinearity generates a nonlocal relaxation
term under renormalization. This means that once a nonlocal
nonlinearity is introduced into the equation, effectively a
certain fractional Laplacian is present as well. This fact can be
observed when applying the SCE method, in the "one-step
renormalization stage" that is part of the solution (see for
example eqs. (6.8)-(6.10) in ref. \cite{SE98}). Then, in addition
to the $q^2 $-relaxation term a $q^{2 - 2\rho } $-relaxation term
appears, and it obviously dominates the dynamics in the
large-scale limit (i.e. $q \to 0$ limit) when $\rho  > 0$. I think
that the failure to realize this fact prevented the other methods
from recovering the exact one-dimensional result.

The last statement can be checked for the methods mentioned above.
My prediction is that if one adds by hand the correct
relaxational term, namely $q^{2-2\rho}$, then at least in one
dimension the right answer will be obtained for the nonlocal
model as well. This idea suggests another important advantage of
SCE, since a straightforward application of SCE yields the extra
fractional Laplacian, and it does not have to be thought of and
added beforehand.

\section{Conclusions}
In this paper several results for the nonlocal Kardar-Parisi-Zhang
equation have been discussed. It has been shown that those
theoretical predictions for the scaling exponents of the NKPZ
equation are inconsistent with a recent one-dimensional exact
result \cite{NKPZ02} that is possible for that model. I also
claimed that these predictions are not compatible with the
expected behavior for the dynamic exponent $z$ as a function of
the nonlocal parameter $\rho$ from general considerations
presented above.

Then, in order to derive reliable results that are consistent with
the exact one-dimensional result an alternative method, namely
the self-consistent expansion (SCE), has been used. The results
obtained using SCE were also consistent with the expected
$\rho$-dependence of the dynamic exponent.  In addition, the SCE
is able to give sensible predictions for higher dimensions, as
well as for any value of $\rho$ and $\sigma$. All the possible
phases found using SCE are summarized in Table I.

I then discuss the implication of this calculation to certain
experimental results \cite{Lei96}-\cite{Zhang92} that might be
well described by a NKPZ model. At the end I also discuss
possible reasons for the failure of all the other methods to
recover the exact one-dimensional result, and to yield sensible
predictions, namely the generation of a fraction Laplacian term
in the equation under renormalization. This observation may
trigger future work on the "fixed" NKPZ model that includes such
a term from the beginning, using methods such as mode-coupling.
It would then be very interesting to compare between the various
results in higher dimensions.

This situation, along with previous results obtained in
\cite{SE92,SE98,katzav99}, suggest that the SCE method is
generally a successful method when dealing with such non-linear
Langevin equations.

 Acknowledgement: I would like to thank Moshe Schwartz for useful discussions.

\newpage

\begin{table}[ht]
\begin{tabular}{|c|c|l|}
\hline $z$ & $\Gamma$ & $Validity$ \\
\hline\hline $2-2\rho$ & $2 - 2\rho  + 2\sigma $ & $\sigma  > \rho
> 0$ and $d > 2 + 2\sigma  - 3\rho $ \\
\hline
 $2-2\rho$ & $2$ & $\rho  > 0$ , $\sigma  < \rho $ and $d >
2
- \rho $) \\
\hline
$2$ & $2 + 2\rho $ & $\sigma  < \rho  < 0$ and $d > 2 + 4\rho $ \\
\hline
 &  & $\rho  < 0$ , $\sigma  > \rho $ and one of
the following: \\
 & & 1. $d > \max \left\{ {2 + 2\sigma  + \rho ,2 + 4\sigma } \right\}$. \\
$2$ &  $2 + 2\sigma $ & 2. $\sigma  > {\textstyle{1 \over 2}}\rho
$ and $d > 2 +
2\sigma + 2\rho $. \\
& & 3. $\sigma  < {\textstyle{1 \over 2}}\rho $ and $2 + 4\sigma < d < 2 + 2\sigma  + \rho $. \\
& & 4. $2 + 2\sigma  + 2\rho  < d < 2 + 4\sigma $. \\

\hline $\frac{{d + 4 - 2\rho  - 2\sigma }}{3}$ &  $\frac{{d + 4 -
2\rho + 4\sigma }}{3}$ & $\sigma  > \rho $ , $d < 2 + 2\sigma  -
\rho  - 3\left| \rho  \right|$ and $d + 4 - 2\rho  + 4\sigma  >
3\Gamma _0
\left( {d,\rho } \right)$ \\

\hline
 $\frac{{d + 4 - \Gamma _0 \left( {d,\rho } \right) - 2\rho
}}{2}$ & Sol. of $F\left( {\Gamma ,z ,\rho } \right) = 0$, & $d <
3\Gamma _0 \left( {d,\rho } \right) - 4 + \min \left\{ {2\rho  -
4\sigma , - 2\rho }
\right\}$ \\
& denoted by $\Gamma _0 \left( {d,\rho } \right)$  &  and $d <
\Gamma _0 \left( {d,\rho } \right) - 2\left| \rho  \right|$
 \\
\hline
\end{tabular}
\caption{A complete description of all the possible phases of the
NKPZ problem, for any value of $d,\rho $ and $\sigma $. The first
two columns give the scaling exponents $z $ and $\Gamma $ for a
particular phase, and the third column states each phase's
validity condition. Note that $\Gamma _0 \left( {d,\rho } \right)$
is the numerical solution of the transcendental equation $F\left(
{\Gamma ,z ,\rho } \right) = 0$ with the scaling relation $z =
\frac{{d + 4 - \Gamma  - 2\rho }}{2}$.} \label{exp_vals}
\end{table}

\newpage
\begin{figure}[htb]
\includegraphics[width=6.5cm]{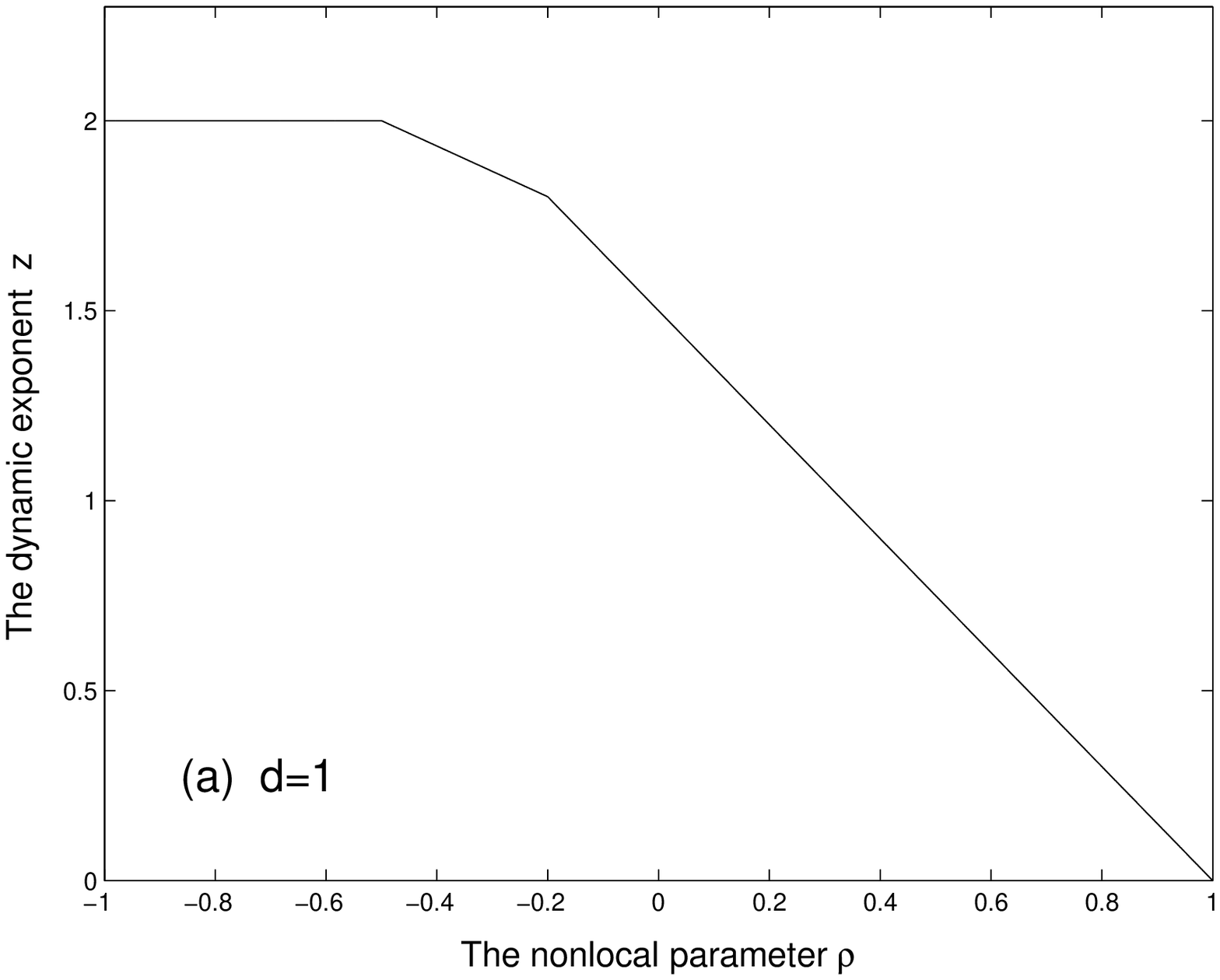}
\label{exps-a}
\end{figure}
\begin{figure}[htb]
\includegraphics[width=6.5cm]{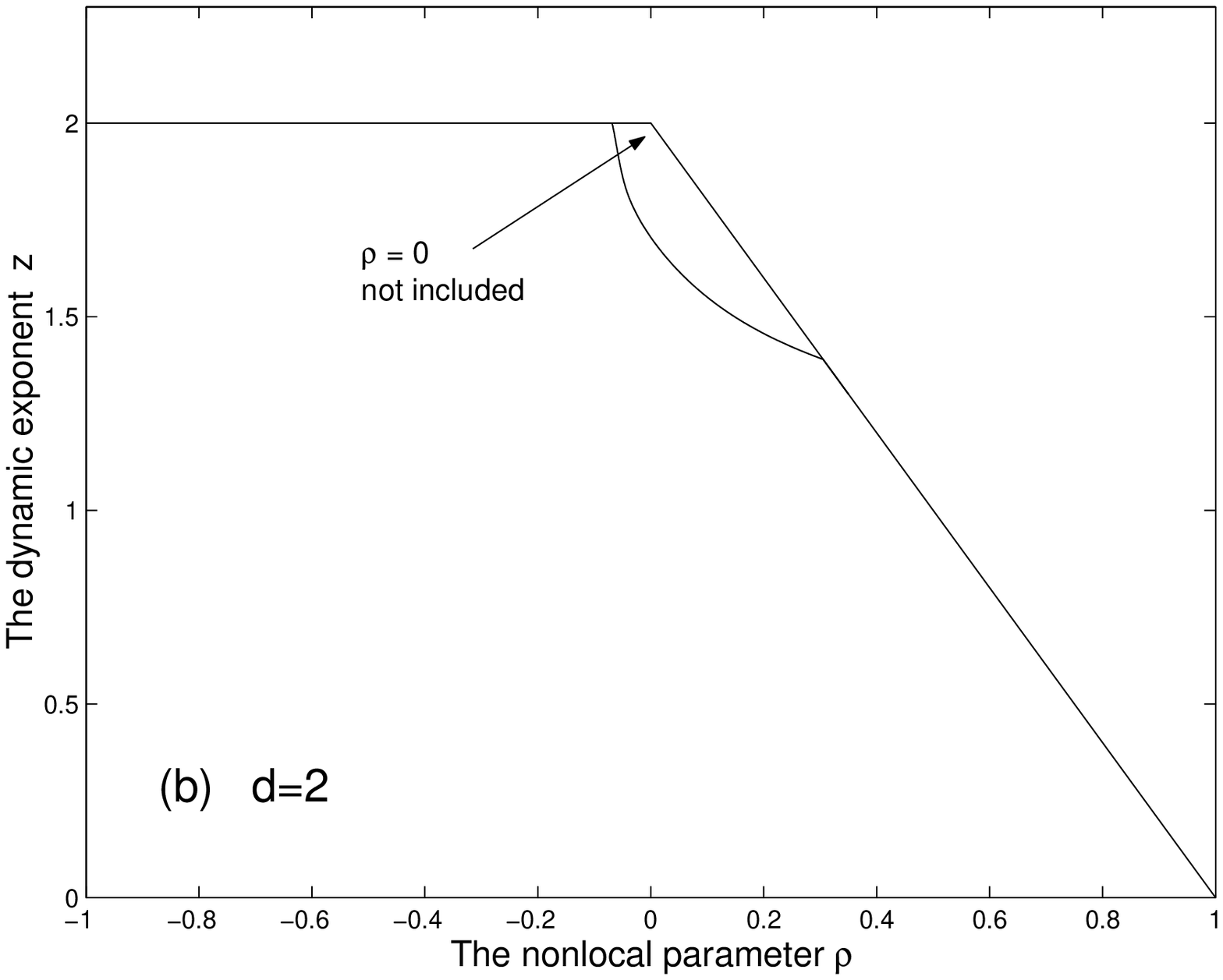}
\label{exps-b}
\end{figure}
\begin{figure}[htb]
\includegraphics[width=6.5cm]{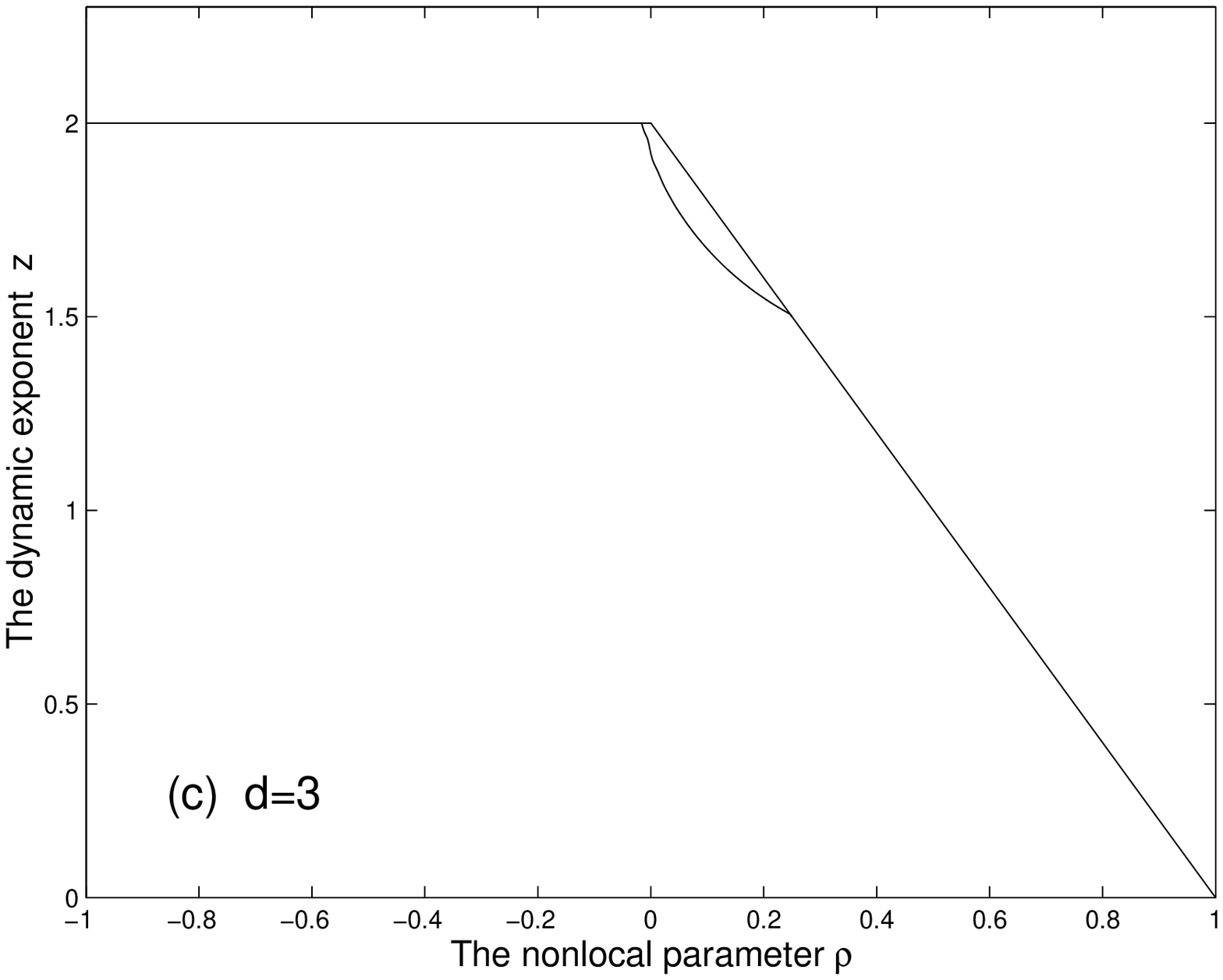}
\caption{The values of the dynamic exponent $z$ as a function of
the nonlocal parameter $\rho $, for uncorrelated noise
($\sigma=0$). (a), (b) and (c) are for $d = 1,2$ and $3$
respectively.} \label{exps-c}
\end{figure}

\end{document}